\documentstyle{l-aa}

\begin{document}

\thesaurus{06
           (06.09.1;   
            06.05.1;   
            02.05.2)}

\title{Solar Models: Influence of Equation of State and Opacity}

\author{M. Y{\i}ld{\i}z \inst{1}, N. K{\i}z{\i}lo\u{g}lu \inst{2}} 
\institute{Ege University, Department of Astronomy and Space Sciences,
           \.Izmir 35100 , Turkey
\and Middle East Technical University, Physics Department, Ankara 06531, Turkey}
 
\date{Received September, 1996; accepted January 22, 1997}

\offprints{M. Y{\i}ld{\i}z}
\maketitle

\begin{abstract}
Solar models through evolutionary phases of
gravitational contraction, pre-main sequence and MS phases, up to
current age $4.5\times10^{9}~yr.$ and
$4.57\times10^{9}~yr.$, were studied adopting different prescriptions
for the equation of state (EOS) and different opacity tables.

Old EOS in Ezer's evolutionary code that we adopted in previous studies
includes pressure ionization, and degeneracy of electrons.
In the present study the EOS is obtained by minimization of the free energy
(Mihalas et al. 1990, MHD), and the radiative opacity is derived from
recent OPAL tables, implemented by the low temperature tables of Alexander
\& Ferguson. The results are compared with solar models we computed with 
different radiative opacities (Cox \& Stewart 1970) and different 
EOS, as with models computed by other authors.

Finally we provide the internal run of the thermodynamic quantities of
our preferred solar model which possesses the following characteristics:
age $ 4.50\times 10^9~yr.$, initial He abundance by mass $0.285$, parameter 
of the mixing length $\alpha=1.82$, radius and 
temperature at the bottom of the convective envelope are $R_b=0.724~R_\odot $
and $T_b=2.14\times 10^6~K$, respectively.\\
~~~~~~~~~~~\\
\end{abstract}
\newcommand{\be}{\begin{equation}}
\newcommand{\ee}{\end{equation}}
\newcommand{\ben}{\begin{eqnarray}}
\newcommand{\een}{\end{eqnarray}}

\section{Introduction}
Recent improvements concerning radiative opacities and equation of state
removed significant discrepancies between observations and theory of
stellar evolution. Updated solar models (Turck-Chi\`{e}ze et al. 1993; Charbonnel \& Lebreton 1993; Gabriel 1994a; Basu \& Thompson 1996) are
in good agreement with the solar helioseismic data.

In this work models are constructed for 
the Sun starting when model becomes stable against gravitational
contraction, with different EOS and opacities,
by using Ezer's stellar evolutionary code (EC) (Ezer \& Cameron 1967; 
Y{\i}ld{\i}z \& K{\i}z{\i}lo\u{g}lu 1995).
EOS of MHD (Mihalas et al. 1990; and references therein)
which uses an approach known as chemical picture
is based on minimization of the free energy and is applicable for stellar 
modeling. We have chosen opacity tables of OPAL (Iglesias et al.  
1992). Since those tables do not extend to
low temperatures, Alexander \& Ferguson opacity (1994) is used for
outer layers of the stars.
The models computed for comparison are obtained using  Cox \& Stewart (1970)
(CS) opacity and  EC EOS which takes
into account degenerate 
electrons and pressure ionization in an artificial way. 
Since OPAL opacity, near the bottom of the convective zone, is larger than the 
CS opacity by a factor of about two, it strongly affects the structure
of the convective zone.
In performing the solar calibration, the lower luminosity due to enhancement of the 
opacity also in radiative regions is balanced by decreasing H (or increasing He)
abundance. Thus, the abundance of H by mass may also be significantly altered.

It was shown that the MHD EOS 
removes significantly the discrepancies between the observational and the 
theoretical frequencies of the solar oscillations (Christensen-Dalsgaard 
et al. 1988). However we do not mention about the solar oscillations.
The incorporation of the Coulomb interaction in EOS reduces both of the pressure
and the energy. To compensate this reduction in pressure, 
the mean molecular weight per free particle, that is He 
abundance, becomes smaller, in contrast to the effect of OPAL opacity. 
The application of MHD EOS requires the calculation of ionization and 
internal energy of each chemical species. Avoiding the 
time consuming calculation of the Saha equation for the heavy elements, we use 
two different methods,
namely, Henyey (Gabriel 1994b) method and Gabriel \& Y{\i}ld{\i}z method (1995).  
 
In Sect. 2 we present the basic features of the code. The EOS with its 
computational method  and opacity used in the construction of the solar models
are given in Sect 3. Sect. 4 is devoted to the influence of MHD  EOS including 
the results of the two methods for the ionization and internal energy of heavy 
elements. In Sect. 5 we give the results of the solar models with different 
EOS and opacity.

\section{Basics of the Modeling}
We solve the four non$-$linear differential equations of the stellar structure
by the method of
Newton-Raphson, from center all the way to the surface, with the surface 
boundary conditions given as
\be
L=4\pi \sigma\,R^2\,T_{eff}^{4},~~~~~~~~~~~~\frac{P_{ph}}{\tau}=
\frac{GM}{R^2\kappa}+\frac{1}{2}a\,T_{eff}^{4}
\ee
where $P_{ph}$ is photospheric pressure, $\tau$ is the optical depth taken to be
$2/3$, and the remaining symbols have the usual meaning.

Mass fraction of heavy elements is taken to be $0.019$.
Chemical abundances of the solar surface found by Anders \&
Grevesse (1989) are chosen. Nuclear reactions are from Caughlan \& Fowler 
(1988). For convection, classical mixing length theory of B\"{o}hm-Vitense 
(1958) and Schwarzschild criteria are used. 

Stellar models are employed mostly in comparative works. Therefore the accuracy
of the code is very important.  For this reason the code is modified to obtain
more accurate models. In order to acquire a higher accuracy,
two points derivative of thermodynamic quantities are replaced  with
quadratic ones. The relative changes of the physical variables become smaller
than $10^{-4}$ for each shell boundaries in the solution of the equation of
stellar structure by the Henyey method.

\section{Input Physics}
\subsection{Opacity}
In order to test the influence of the uncertainities in the radiative opacities
we have used different sets of the opacity.
The OPAL opacity is available for the temperatures larger than $6000~K$,
and implemented by the Alexander \& Ferguson opacity table
at low temperatures (OA). 
For the Sun, the transition point ($T=\,8000$  K) 
is chosen such that it always belongs to convective region in all 
evolutionary phases of the Sun. So, small discrepancies may not
cause any problem in the solution of finite-difference equations. 
We calculate the opacity by quadratic interpolation in chemical
composition, $T$ and $R~(R=\rho/T_6^3$).

Model 1 and 2 are obtained using the radiative opacity of CS.
The opacity tables include three H-He mixtures in which H/He =$4$; $1$; 0.
Required opacity for the solar material is found by logarithmic interpolation 
between the three sets of the opacity tables. 
\subsection{Equation of State: Minimization of Free Energy}
The free energy is minimized in 
order to obtain the Saha equation which is required to find the number of 
particles in each degree of ionization.
  This is the chemical picture in which atoms and molecules are described as separate
entities.
 
Since Fermi and Bose gases obey different statistics, their free energies,
$F$, are 
different. But, disregarding signs, they have the same form:   
\be
 F =\Omega + \mu N, {\label{c2.9}}
\ee
\be
 \Omega = -PV = \mp kT \sum_{i} \ln (1\pm e^{(\mu - \epsilon_{i})/kT}),
  {\label{c2.10}}
\ee
where summation is over all quantum states $i$, $\mu$ is the chemical
potential, $\epsilon_{i}$ is the energy
of this state, and $k$ is the Boltzman constant. For Fermi (Bose) gas upper 
(lower) sign in Eq. (\ref{c2.10}) applies. Total number of particles,
$N$, is given by the summation of the average occupation number over all 
quantum states, $n_i$.
\subsubsection{Degeneracy of Electrons} 
The free energy of a gas of particles with half integral spin is not as simple 
as that of the photon gas in those regions of the  $\rho-T$ plane where the
gas is partially degenerate and partially relativistic. 
In the region of interest, there is no analytic solutions of the EOS
except at some
points in which both degeneracy parameter $\lambda $ ($\lambda=\mu/kT$)
and relativistic parameter $\beta $ ($\beta = kT/mc^2$) have extreme values.
Therefore
numerical methods should be employed for a semi-relativistic and
semi-degenerate
gas. As $Q_{n}$ functions introduced by Guess (1966) are better suited
for numerical calculations than Fermi-Dirac functions, we express
$\Omega_e$ as
\be
\Omega_{e} = -\frac{\pi}{3}\,\frac{m^4c^5}{h^3}\,V\left(Q_4-4Q_2+3Q_0\right).
 {\label{c2.31}}
\ee
Like $\Omega_{e}$, the number density of electrons $n_e$
is an implicit function of $\lambda $ and $\beta $.  In terms of
the functions $ Q_{n}$  it is given as
\be
 n_{e}=\frac{N_{e}}{V}=2\pi\,\left(\frac{mc}{h}\right)^{3}\left(Q_3-Q_1\right),
  {\label{c2.33}}
\ee
where $N_e$ is the number of electrons in volume $V$. If $\lambda\;>>\; 0$
then perfect gas relation is recovered and $\lambda$ for the non-relativistic 
region can be given in terms of $n_e$ and $T$:
\be
e^{-\lambda}=\frac{h^3}{2(2\pi m k)^{3/2}}\;\frac{n_e}{T^{3/2}},{\label{c2.33a}}
\ee
which corresponds to slightly degenerate or non-degenerate case.

Then the energy and the pressure are  given by the derivatives of the
free energy with respect to the temperature and to the volume,
respectively:
\be
E_{e}= \pi \,\frac{m^4c^5~V}{h^3}\,\left(Q_4+Q_0\right) ,  {\label{c2.40}}
\ee

\be
P_{e}=\frac{\pi}{3}\frac{m^4c^5}{h^3}\,\left(Q_4-4Q_2+3Q_0\right). 
{\label{c2.41}}
\ee
\subsubsection{Boltzmann Gas and Partition Function}
Free energy for the Boltzmann gas can be written as follows
\be 
F_{b}=\sum_{i,j} F_{bi}^{j}= -kT\sum_{i,j} N_{i}^{j}
\ln \left[ \frac{eV}{N_{i}^{j}}
\left(\frac{2\pi m_{i}kT}{h^2}\right)^{3/2} B_{i}^{j}\right].  {\label{c2.50e}}
\ee
where $N_{i}^{j}$ and $B_{i}^{j}$ are number of $i$-type $j$-times
ionized ions in volume $V$ and partition function of these ions,
respectively.

Now, energy  and pressure can be found for a Boltzman gas by 
differentiating $F_{b}$ given above. Since partition function is also a
function of temperature and density (due to interaction between
particles), its derivatives must also be taken into account:  
\be
E_{b}= 
  kT\sum_{i,j} N_{i}^{j}\left(\frac{3}{2}-\sum_{m=j+1}^{Z}\frac{I_{i}^{m}}{kT}
+\frac{\partial \ln U_{i}^{j}}{\partial \ln T} \right), {\label{c2.50f}}
\ee
and
\be
P_{b}=
kT\sum_{i,j} n_{i}^{j}\left(1-\frac{\partial{\ln
U_{i}^{j}}}{\partial{\ln \rho}}\right)  {\label{c2.50g}}
\ee
where $n_{i}^{j}\equiv N_{i}^{j}/V$ and
$B_{i}^{j}=e^{\epsilon_{i,0}^{j}/kT} U_{i}^{j}$. The second quantity in
the parenthesis in Eq.({\ref{c2.50f}}) is $\epsilon_{i,0}^{j}/kT$ where
$I_i^m$ is the $m$th ionization potential of $i$-type ion 

Influence of interaction between atoms (ions) may obliterate
some outer states of an atom. Electrons occupying these states are not
bounded to the atom any more. This is known as pressure ionization. MHD
have shown how to include the effect of pressure ionization into the
partition function based on \"{U}nsold's theory (Clayton 1968).
According to  \"{U}nsold's theory partition function can be
written as
\be
U_i^j=\sum _{k}g_{i,k}^j w_{i,k}^j e^{-\chi_{i,k}^j/kT}~~.
{\label{c2.53a}}
\ee
where $w_{i,k}^j$ is the survival probability of the state k for
a $j$-times ionized $i$-type atom with $j$ less electron,
$g_{i,k}^j$ and $\chi_{i,k}^j$ are 
the statistical weight and excitation energy of this state,
respectively.

If the survival probability for any state is zero, neither this state
nor higher states contribute to the
partition function. Apart from  thermodynamical consistency the
survival probability given by MHD  has important advantages.
The partition function doesn't explode, and transition from bound to
free
states is continuous since $w_{i,k}^j$ is continuous.
\subsubsection{Coulomb Interaction}
At relatively high densities atoms are very close to each
other. A bound electron may not preserve
its original state under the influence of the Coulomb potential due
to other charged particles in the plasma. The  Coulomb
interaction implies that slight changes occur in the energy and the pressure.
An approximate expression for the Coulomb energy is given by Landau \&
Lifshitz (1969) as
\be
E_{c}=- q_{e}^{3} \left(\frac{\pi}{kTV}\right)^{1/2}
\left(\sum_{i,j}N^{j}_{i}j^2\right)^{3/2}
= -\frac{1}{2} q_e^2 \frac{\sum_{i,j}N^{j}_{i}j^2}{R_D} {\label{c2.51}}
\ee
where $q_e$ is proton charge and $R_D$ is Debye radius.
Expression for pressure due to interaction  is
\be
P_{c}
=-\frac{q_e^3}{3V} \left(\frac{\pi }{kTV}\right)^{1/2}
\left(\sum_{i,j}N^{j}_{i}j^2\right)^{3/2}. {\label{c2.53}}
\ee
As seen in Eqs. ({\ref{c2.51}}) and ({\ref{c2.53}}) both $E_{c}$ and $P_{c}$ 
are negative, since the Coulomb interaction causes a decrease in the energy and
the pressure.
\subsubsection{Saha Equation } 
To find equilibrium number of particles it is necessary to
minimize the free energy using stoichiometric relations describing
ionization (or dissociation) processes.
Then the Saha equation is obtained with ionization potential
$I_{k}^{l}$
of $l$-times ionized $k$-type atom:
\ben
\frac{N_{k}^{l+1}}{N_{k}^{l} }&=&
\frac{U_{k}^{l+1}}{U_{k}^{l} }
\exp\biggl\{\lambda -\bigl\{I_{k}^{l}-2(l+1) c_2 
(N {\overline {Z_{eff}^{2}}})^{1/2}\bigr\}/kT \nonumber \\
&+&\sum_{i,j} N_{i}^j\left(\frac{\partial \ln U_i^j} 
{\partial N_k^{l+1}}- \frac{\partial \ln U_i^j}
{\partial N_k^{l}}\right)\biggr\} ~~~~. 
{\label{c2.72}}
\een
where
\be
{\overline {Z_{eff}^{2}}}= \frac{1}{N}\sum_{i,j}N^{j}_{i}j^2
{\label{c2.71a}}
\ee
with N being the total number of particles in the system including
electrons.

Two remarks are called for Eq. (\ref{c2.72}) which includes 
degeneracy of electrons and the Coulomb interaction. First, any  
additional term to the free energy appears with its
partial derivative with respect to the number density of particles and
as an argument of the exponential in the Saha equation.
Second, both of the terms coming from the free energies  of the degenerate
electrons and the Coulomb
interaction behave like  processes lowering ionization potential.
Therefore, some of the early works on EOS (Harris 1964; Graboske et al. 1969)
treated ionization potential somehow 
as a function of the thermodynamical variables.

\subsection{Computational Method of EOS}  
The total energy and pressure of the plasma are the summation of the
partial contributions  of the effects discussed in the previous sections
and the radiation term. 
Most of the physical quantities appearing in the equations for
energy and pressure of different kinds of gases can be identified, explicitly
or implicitly, in terms of number density of electrons. Therefore the
EOS is calculated by iterating over the number of moles of electrons per unit 
mass, $N_e$ (Gabriel 1994b). Then the problem reduces to solve the
Saha equation for any given value of $N_e$.

We implemented the whole procedure in our stellar evolution code in the
following way. Starting from the center, an initial guess of $N_e$ is
obtained by assuming the complete ionization,
 that is each species has charge $Z_i q_e$, which is sufficiently accurate 
for the center of a star. And then, in
outer shells, the first guess of $N_e$ is taken as its last computed 
value in the previous shell, until the surface is reached.

From the value of $N_e$,  the degeneracy parameter $\Lambda=e^{-\lambda}$ is 
found from Eq. (\ref{c2.33a}). 
In the transition region $0.0005 < \Lambda < 600.0$,
gas is partially degenerate and $\Lambda$ can be recalculated from the  
fitting formula of Henyey (Gabriel 1994b; Y{\i}ld{\i}z 1996).

Because of the low abundance of heavy elements in the solar mixture, probably it is
unnecessary for those elements, to calculate the quantities such as 
partition functions and the number of elements at any degree of ionization, 
with high accuracy. Only for H and He very precise calculations need
to be done. 

In order to save computer time, all that is needed is to calculate the
effective
 degree of ionization (i.e., effective charge of each element)
 \be
 Z_{eff,i}=\displaystyle \frac{\sum_{j} N_i^j j}{N_i}~~~~.
{\label{c2.84}}
 \ee
 Then, there is no need to solve the Saha equation.
Two methods are used to compute the effective charges
of  $^{12}$C, $^{14}$N, $^{16}$O,
 $^{20}$Ne and the binding energies of these atoms (ions). 
The first one is Henyey's fitting method (Gabriel 1994b; Y{\i}ld{\i}z 1996) 
which uses $N_ekT$ and the chemical potential of electron $\mu_{e}$ as input
parameters. The second method is due to  Gabriel \& Y{\i}ld{\i}z (1995),
which is similar to Henyey's method but more precise. It takes into account
Debye radius as an additional parameter.

Since heavy elements are rare in the atmospheres of the stars
only the molecules of hydrogen, H$_2$, H$_2^+$, and H$^-$ are 
considered. The abundance and the energies of these molecules are
solved by the method of Vardya (1961).

\section{Influence of MHD EOS} 
In order to compare the results and the influence of MHD EOS, the 
required density, temperature and chemical composition are taken from a
model of the present Sun (Model A, see Table 1).
Henyey method is used for ionization of heavy elements, unless stated otherwise.

   \begin{figure}[htbp]
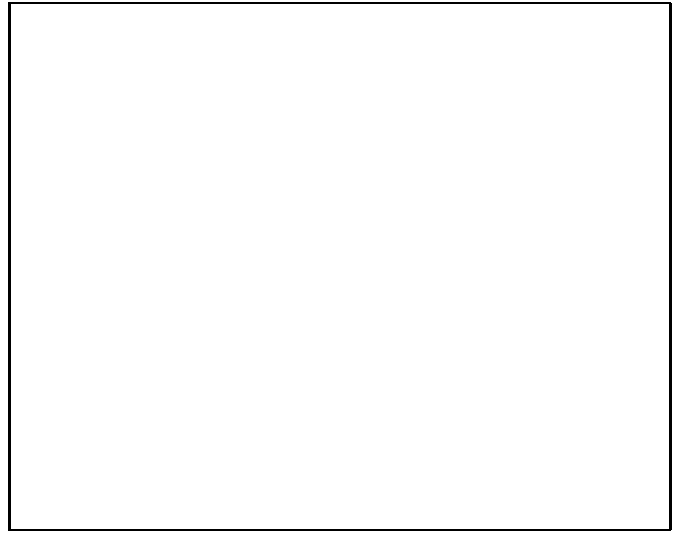

     \picplace{7cm}
      \caption{The fractional contribution to
the pressure due to Coulomb interaction, degeneracy of electrons and
derivative of the partition functions of H, He, and He$ ^+$ with respect
to density, is plotted as a function of the logarithm of the temperature for a
selected solar model (case A below).
              }
   \end{figure}

   \begin{figure}[htbp]
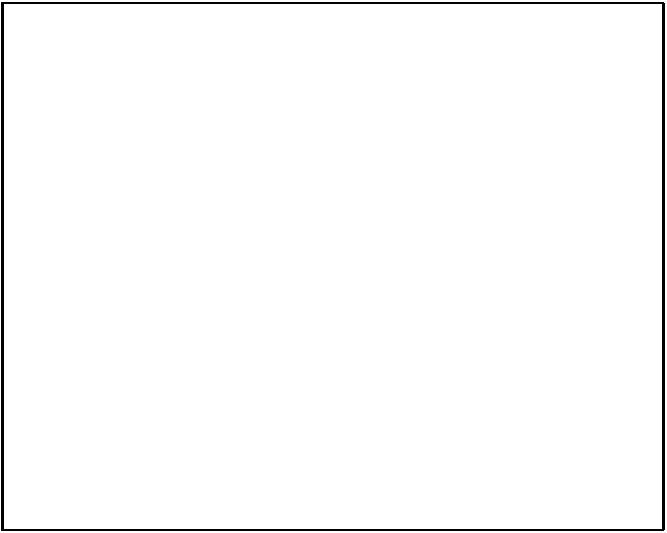

     \picplace{7cm}
\caption{ The run of the effective charge of several elements as a function
          of the logarithm of the temperature in the model of Fig. 1.
          The Saha equation is solved for H and He, and
          Henyey's method is used for heavy elements.  }
   \end{figure}

   \begin{figure}[htbp]
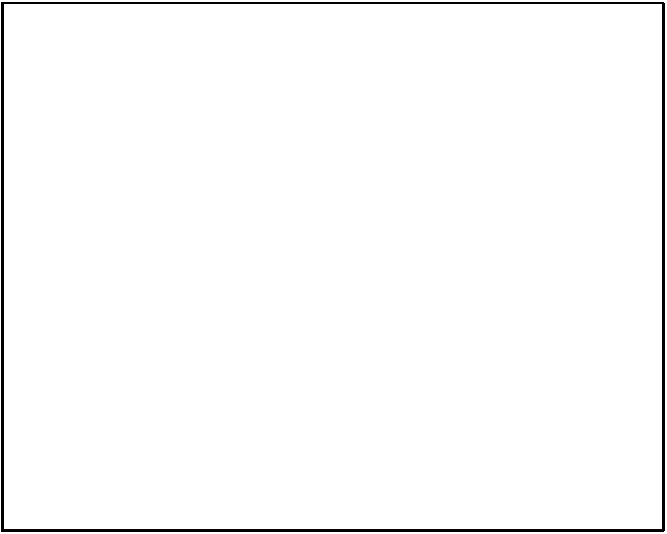

     \picplace{7cm}
      \caption{
 The run of the effective charge of several elements as a function of the
logarithm of the temperature in the model of Fig. 1.
The Saha equation is solved for H and He, and
the method of Gabriel \& Y{\i}ld{\i}z (1995) is used for heavy elements.
              }
   \end{figure}

   \begin{figure}[htbp]
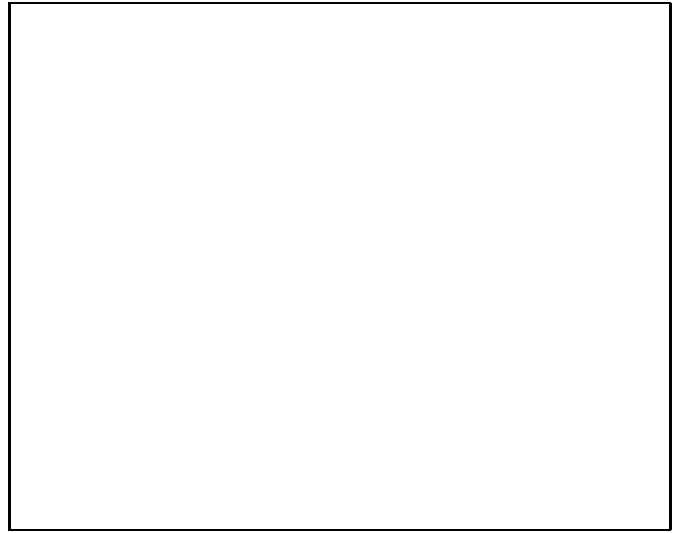

     \picplace{7cm}
      \caption{
The ratio of adiabatic gradients resulting from MHD EOS with Gabriel
\& Y{\i}ld{\i}z (1995) (solid line) and with Henyey (dots) methods for
ionization of heavier elements, to that obtained adopting the EOS of old
routine in Ezer's code (EC) for the selected model of Fig. 1.
              }
   \end{figure}

Contribution of the Coulomb interaction,  the degeneracy of the electrons and derivative  
of the partition function to the total pressure can be seen in Fig.1.
 For the Model A $P/(P-P')$ is plotted with respect to
$~\log(T)$ of temperature  from surface to the center
throughout the solar model, where $P$ is the total pressure, and $P'$ is 
any of fractional contribution to the pressure.
Thin solid line is for the Coulomb pressure. 
The ratio is less than or equal to one, since the Coulomb term causes a decrease
in pressure. At the surface the Coulomb pressure is zero.
 As the temperature and the density, and therefore 
the degree of ionization increase, fractional change becomes a few percent. 
Contribution of the Coulomb potential to the total
pressure is maximum ($6\%$) between $~\log (T)=~4.6$ and $~4.7$, at which point
the most abundant species (hydrogen and 
helium) are completely and singly ionized, respectively. This means that the 
Debye radius is sufficiently small. As the temperature and the density increase
the contribution becomes smaller, until partial degeneracy of the electrons sets in
near the center of the Sun.

The pressure of the degenerate-electrons calculated by Henyey fitting
formula (also calculated by method of Y{\i}ld{\i}z \& 
Eryurt-Ezer (1992) using
Guess functions, the difference is less than $0.01 \%$), represented by dots,
deviates from one at the central part. Here $N_e kT $
is subtracted from the total electron pressure.
At the center of the Sun degeneracy parameter ($\lambda$) is equal to 
$ ~-1.559~$ which implies a slight degeneracy, and the contribution 
of degenerate-electrons is about $2\%$.

Thick solid line in Fig.1. is a superposition of three Gaussian
curves. From left to right, the first one is dominantly the density derivative
of hydrogen partition 
function which is maximum ($\approx~ 0.5\%$) at $~\log (T)\approx ~4.5$. 
Second one is
due to the density derivative of He partition function, which is maximum at
about $~\log (T)\approx ~4.75$ and its contribution to the total pressure
is less than $1\%$.
Last one is
due to the density derivative of He$ ^+$ partition function, which is non-zero 
in an important part of the solar interior but with a lower contribution to the
total pressure than
the others. The base temperature of the convective zone in the Model A is
about $2.14\times 10^6~K$. This corresponds to a point $(\log(T)=6.3$)
at which He$^+$ starts to appear.

The Saha equation, Eq ({\ref{c2.72}}), is solved exactly for H and He. 
Results obtained from Henyey method, the effective charges divided by the atomic numbers of each 
element, throughout the Model A, are given in Fig.2. Due to its low ionization 
potential hydrogen (solid line with asterisks) is rapidly completely ionized 
(at $~\log (T)\approx~5.2$). For the effective 
charge of He (solid line with circle) two phases are seen.
Even at the center, none of the heavier elements, C (solid line), 
N (dashed line), O (solid line with triangle), Ne (dashed line with 
asterisks) are completely ionized. Owing to rough calculation, each
line of the heavy elements cross each other near
the center. On the other hand, at the center the radiation and the gas
pressure (and energy) are sufficiently large, therefore 
the computed values are within an acceptable range.
This is not the case at the surface where binding energy has an important role 
in the total energy. While H and He are neutral, each of the heavier elements is in the
state of first ionization. 
\begin{table*}[t]
\begin{center}
\caption{ Comparison of the present solar models with different EOS,
opacities and data of the Sun.}
{\begin{tabular}{|l|cccccllcc|}       \hline
Models   & Y    &$\alpha$&$R_{c}/R_{\odot}$&$T_{b}/10^6$&$T_{eff}$&$T_c/10^6$&$\rho_
{c}$
&$X_c$& $P_c/10^{17}$\\
\hline
1 (CS EC D1)       & 0.224& 1.32 &0.19 & 1.24& 5804 & 15.00& 136.4&0.438 &2.244\\
2 (CS MHD+H D1)    & 0.205& 1.46 &0.20 & 1.31& 5794 & 14.91& 132.0&0.478 &2.147\\
3 (OA EC D1)       & 0.302& 1.57 &0.27 & 2.07& 5799 & 15.80& 151.9&0.324 &2.342\\
4 (OA MHD+H D1)    & 0.284& 1.76 &0.27 & 2.11& 5797 & 15.70& 149.3&0.338 &2.271\\
A (OA MHD+GY D1)   & 0.285& 1.82 &0.276& 2.14& 5799 & 15.70& 147.1&0.345 &2.269\\
B (OA MHD+H  D1)   & 0.285& 1.78 &0.277& 2.15& 5799 & 15.69& 147.5&0.343 &2.267\\
C (OA MHD+GY D2)   & 0.283& 1.82 &0.276& 2.16& 5792 & 15.67& 148.6&0.339 &2.270\\
\hline
CL (1993)                & 0.278& 1.66 &0.281& 2.13&      & 15.56&150.7& &    \\
BU (1988)                & 0.271&      &0.277& 2.11&      & 15.62&148  &0.341 &2.290
    \\
TCCD (1988)              & 0.275& 2.11 &0.270& 2.04&      & 15.52&147.2 & &\\
CGK (1989)               & 0.291& 1.89 &0.286& 2.29& 5770 & 15.68&162.4  &0.352 &2.2
78\\
SBF (1990)               & 0.278& 2.07 &0.260& 1.96&      & 15.42&146.6  & & \\
LD (1988)                & 0.278& 2.16 &0.265&2.02 & 5781 & 15.54&148.0 &0.352&2.278
  \\
CDR (1996)               & 0.265& 1.77 &0.274&2.10 &      & 15.45&147.2 &0.372&  \\
\hline
\end{tabular}}
\end{center}
\end{table*}

A physically more reliable method is that of Gabriel \& Y{\i}ld{\i}z
(1995). As seen 
in Fig.3, at the middle of the Sun, ionization degrees of the heavier elements 
are a little bit smaller than, but comparable to those given in Fig.2. At the 
center and at the surface, the values of ionization degree are not in 
contradiction with the basic truths of atomic physics.
All atoms are neutral at the surface. Ionization of C starts before H, and 
ionization of N, O,and Ne starts after H but before He, as expected
from the atomic physics.

The effect of the MHD EOS on the adiabatic gradient is shown in  Fig.4. There
we plot the ratio of the adiabatic gradient resulting from the  
Gabriel \& Y{\i}ld{\i}z (solid line) and from  the Henyey method (dotted
line) to the value obtained with EC EOS are given.
Both are larger than the old gradient, and three peaks correspond
ionization zones of H, He and He$^+$.

\begin{table*}[t]
\begin{center}
\caption{ The present Sun model (Model A) with OPAL opacity and MHD EOS
by using method of Gabriel and Y{\i}ld{\i}z for ionization of heavy elements}
\begin{tabular}{|r|cccccrcc|}       \hline
Sh.n &$M/M_\odot$&$R/R_\odot$&$\rho$&$T$&  $P$     & $E~~~~~~$ & $L/L_\odot$& H\\
\hline
  1 &0.0000&0.000& .1471E+3 & .1570E+8 & .2269E+18 & .2289E+16 &0.000 & .3455  \\
  7 &0.0004&0.016& .1446E+3 & .1560E+8 & .2232E+18 & .2290E+16 &0.004 & .3518  \\
 17 &0.0017&0.019& .1436E+3 & .1557E+8 & .2217E+18 & .2291E+16 &0.007 & .3543  \\
 41 &0.0022&0.028& .1397E+3 & .1542E+8 & .2158E+18 & .2292E+16 &0.021 & .3640  \\
 66 &0.0074&0.042& .1309E+3 & .1510E+8 & .2028E+18 & .2301E+16 &0.066 & .3889  \\
106 &0.0550&0.089& .9411E+2 & .1355E+8 & .1470E+18 & .2321E+16 &0.368 & .5126  \\
126 &0.2019&0.155& .5372E+2 & .1100E+8 & .7553E+17 & .2089E+16 &0.804 & .6373  \\
133 &0.2807&0.182& .4166E+2 & .1002E+8 & .5442E+17 & .1941E+16 &0.901 & .6636  \\
155 &0.5582&0.279& .1539E+2 & .7289E+7 & .1490E+17 & .1438E+16 &0.996 & .6921  \\
179 &0.7825&0.393& .4271E+1 & .5241E+7 & .2972E+16 & .1036E+16 &1.000 & .6959  \\
219 &0.9424&0.593& .5375E+0 & .3195E+7 & .2281E+15 & .6330E+15 &1.000 & .6962  \\
236 &0.9701&0.683& .2346E+0 & .2511E+7 & .7823E+14 & .4977E+15 &1.000 & .6962  \\
256 &0.9787&0.724& .1659E+0 & .2144E+7 & .4720E+14 & .4245E+15 &1.000 & .6962  \\
280 &0.9912&0.808& .8149E$-$1 & .1340E+7 & .1444E+14 & .2634E+15 &1.000 & .6962  \\
298 &0.9958&0.857& .4752E$-$1 & .9381E+6 & .5878E+13 & .1831E+15 &1.000 & .6962  \\
356 &0.9994&0.933& .1265E$-$1 & .3927E+6 & .6457E+12 & .7441E+14 &1.000 & .6962  \\
369 &0.9996&0.945& .9133E$-$2 & .3174E+6 & .3747E+12 & .5947E+14 &1.000 & .6962  \\
389 &0.9998&0.961& .5046E$-$2 & .2167E+6 & .1394E+12 & .3937E+14 &1.000 & .6962  \\
409 &0.9999&0.970& .3053E$-$2 & .1589E+6 & .6090E+11 & .2753E+14 &1.000 & .6962  \\
425 &0.9999&0.976& .1993E$-$2 & .1246E+6 & .3067E+11 & .2005E+14 &1.000 & .6962  \\
442 &1.0000&0.981& .1345E$-$2 & .1012E+6 & .1656E+11 & .1463E+14 &1.000 & .6962  \\
458 &1.0000&0.985& .8592E$-$3 & .8004E+5 & .8222E+10 & .9747E+13 &1.000 & .6962  \\
475 &1.0000&0.987& .5616E$-$3 & .6345E+5 & .4179E+10 & .6144E+13 &1.000 & .6962  \\
493 &1.0000&0.990& .3518E$-$3 & .4986E+5 & .2000E+10 & .3096E+13 &1.000 & .6962  \\
512 &1.0000&0.992& .2014E$-$3 & .3948E+5 & .8721E+09 & .3266E+12 &1.000 & .6962  \\
522 &1.0000&0.993& .1454E$-$3 & .3527E+5 & .5502E+09 &$-$.9956E+12 &1.000 & .6962  \\
533 &1.0000&0.994& .1043E$-$3 & .3179E+5 & .3475E+09 &$-$.2183E+13 &1.000 & .6962  \\
543 &1.0000&0.995& .7647E$-$4 & .2910E+5 & .2282E+09 &$-$.3169E+13 &1.000 & .6962  \\
563 &1.0000&0.996& .3947E$-$4 & .2464E+5 & .9505E+08 &$-$.4962E+13 &1.000 & .6962  \\
583 &1.0000&0.997& .2024E$-$4 & .2139E+5 & .4031E+08 &$-$.6460E+13 &1.000 & .6962  \\
594 &1.0000&0.997& .1433E$-$4 & .2002E+5 & .2607E+08 &$-$.7143E+13 &1.000 & .6962  \\
603 &1.0000&0.998& .1045E$-$4 & .1892E+5 & .1756E+08 &$-$.7723E+13 &1.000 & .6962  \\
613 &1.0000&0.998& .7633E$-$5 & .1792E+5 & .1188E+08 &$-$.8267E+13 &1.000 & .6962  \\
623 &1.0000&0.998& .5036E$-$5 & .1674E+5 & .7103E+07 &$-$.8944E+13 &1.000 & .6962  \\
643 &1.0000&0.999& .2610E$-$5 & .1508E+5 & .3158E+07 &$-$.9943E+13 &1.000 & .6962  \\
663 &1.0000&0.999& .1296E$-$5 & .1348E+5 & .1325E+07 &$-$.1096E+14 &1.000 & .6962  \\
674 &1.0000&1.000& .8965E$-$6 & .1266E+5 & .8333E+06 &$-$.1150E+14 &1.000 & .6962  \\
694 &1.0000&1.000& .4730E$-$6 & .1108E+5 & .3607E+06 &$-$.1250E+14 &1.000 & .6962  \\
705 &1.0000&1.000& .3356E$-$6 & .9903E+4 & .2195E+06 &$-$.1311E+14 &1.000 & .6962  \\
715 &1.0000&1.000& .2679E$-$6 & .8210E+4 & .1407E+06 &$-$.1363E+14 &1.000 & .6962  \\
716 &1.0000&1.000& .2658E$-$6 & .8030E+4 & .1363E+06 &$-$.1366E+14 &1.000 & .6962  \\
717 &1.0000&1.000& .2642E$-$6 & .7829E+4 & .1320E+06 &$-$.1370E+14 &1.000 & .6962  \\
727 &1.0000&1.000& .2421E$-$6 & .6432E+4 & .9894E+05 &$-$.1387E+14 &1.000 & .6962  \\
739 &1.0000&1.000& .1682E$-$6 & .5799E+4 & .6197E+05 &$-$.1393E+14 &1.000 & .6962  \\
\hline
\end{tabular}
\end{center}
\end{table*}

\section{Results and Discussion}
Recently we have gathered very important information about the internal 
structure of the Sun by helioseismological investigations.
Distance of the convective zone from the center of 
the Sun is determined from $p$-mode oscillations as $0.713\pm0.003~R_\odot$  
by Christensen-Dalsgaard et al. (1991).
Photospheric abundance of He by mass is found 
as $0.242\pm 0.003$ (Hernandez \& Christensen-Dalsgaard 1994).
There are several attempts to determine the central density of the Sun
from helioseismological data. Results obtained by
Dziembowski et al. (1994) are in between $130-150~g~cm^{-3}$, while Gough \&
Kosovichev (1988) quote $\approx 160~g~cm^{-3}$ and Vorontsov \& Shibahashi 
(1991) quote a range of $110-115~g~cm^{-3}$.

Two sets of the solar data are used in the calculations.
In one set (D1), the age of the Sun is taken
as $4.5\times 10^9~yr.$, and its 
luminosity and radius are 
$3.90\times 10^{33}~erg~s^{-1}$
and $6.951\times 10^{10}~cm$, respectively.
In the other set (D2) (case C below), the luminosity and the  age of the Sun 
are taken from Bahcall et al. (1995), namely
$L_\odot=3.8440\times 10^{33}~~erg~s^{-1}$ and $~~~~t_\odot=4.57\times
10^9~~yr$.  The radius of the Sun 
 is $R_\odot=6.9598\times 10^{10}~cm$
(Sackmann et al. 1993). In both of the sets the
solar mass is $1.985\times 10^{33}~g$.
 
For different combinations of EOS's and opacities and for different solar data,
starting from threshold of stability point at which gravitational and internal 
energies are nearly the same, we constructed a series of evolutionary models 
for the Sun. Each model given in Table 1 is obtained
by changing initial mass fraction of hydrogen, helium  and convective parameter
$\alpha$, in order to fit the Sun's luminosity and radius to the present data,
with different accuracies. 
The first column in Table 1 represents 
the initial abundance of He by mass, and $\alpha$ is
given in the second column.  The depth of the 
convective zone in terms of solar radius, $R_c/R_{\odot}$, is shown in the
third column, and base temperature of the convective zone, $T_b$, is 
in the fourth column. Last five
columns are the effective temperature ($T_{eff}$), the central temperature
($T_c$), density($\rho_c$), abundance of hydrogen by mass at the center ($X_c$),
and the central pressure ($P_c$), respectively (in $cgs$).
All the models, except Model C, is with D1.
The first three models are constructed with an accuracy of a few percent,
and the fourth model with an accuracy of $0.5\%$. Models A, B and C are more
precise models (accuracy is $10^{-4}$). The difference between the Model A
and the Model B is the method of heavy element ionization. The latter (also
Model 4) uses Henyey's
method (MHD+H), the former is obtained by the method of Gabriel \& Y{\i}ld{\i}z
(MHD+GY) as the Model C.

When we compare each of the first four models with the model of the
same EOS but different opacity (Model 1 with 3, and Model 2 with 4), it is seen
that OA opacities strongly influence the structure of the Sun. 
The enhancement of the opacity (OPAL) below the convective 
zone enlarges the zone's size toward the center of the Sun. 
When  MHD is used, OA opacities enlarge the convective zone by about $40 \%$, 
and by $35 \%$ if EC EOS is used.
Increase in base temperature of convective zone is about
$61 \%$ with MHD EOS and $67 \%$ with EC EOS whenever OA opacities are
employed in place of CS opacity. 
While He abundance by mass
is exceeding the helioseismological result, the central density is about 
$150~g~cm^{-3}$, and the base temperature of the convective zone is larger than
$2\times 10^6~K$ in the models with OA. 
Because of  increase  in opacity, less energy reaches the surface. 
This is  compensated by an increase in He abundance. The convective parameter
alpha increases by about $30\%$, when OA is employed in place of CS.

If we compare Model 1 with 2, and Model 3 with 4, it becomes obvious that
the depth and the base temperature of the convective zone is not very sensitive
to the EOS. The values of He abundance and $\alpha$ are changed by EOS. 
Incorporation of the Coulomb interaction reduces the  energy and pressure by 
some fraction. In order to compensate this reduction in pressure, mass fraction
of He (or molecular weight per free particle) decreases by about $6\%$
in case of CS, and $10\%$ in case of OA opacities. The only difference 
between the ways that Model A (Gabriel \& Y{\i}ld{\i}z) and B (Henyey) are
obtained, is the calculation method of the ionization of heavy elements 
and their internal energies. There are small differences between these models.
The convective parameter $\alpha$ of Model A is a little bit larger than that 
of Model B, since the pressure scale height (the number density of electrons) 
of Model A in outer layers is less than that of Model B.

The small differences between Models B and 4 are due to low accuracy of Model 4.
The Models A (D1) and C (D2) have different solar data. The Model C has
lower luminosity and higher age than the Model A.

In Table 1,
we also give models constructed by Charbonel \& Lebreton (CL) (1993), Bahcall 
\& Ulrich (BU) (1988), Turck-Chieze et al. (TCCD) (1988), Cox et al. (CGK)
(1989) , Sackmann et al. (SBF) (1990), Lebreton \& D\"{a}ppen (LD) (1988),
and Ciacio et al. (CDR)(1996).
Similar to our models, all the models obtained by different authors are 
standard models, that is there is no rotation, no diffusion process and
magnetic field is negligible.
Except model of CGK, the initial mass fraction of He in Models A, B and C is a
little bit larger than those found by other researchers. 
The extreme model of CGK, which has the largest value of He mass fraction, base 
temperature of the convective zone, and the central temperature,
has an age of $4.66\times 10^9~yr$. 
The EOS of CGK is very similar to EC EOS,
and Iben's fitting formula (1975) is used for opacity.  
The model of CDR which has the same solar data as Model C except the
solar mass and heavy element abundance, is obtained by using both the EOS 
and the opacity of OPAL. The significant difference between these two
models is in He abundance due to their data of high solar mass and low
metallicity. 
The other models take the age of the Sun as $4.6\times 10^9~yr$. Therefore, 
their He mass fractions are closer to, but less than, that of our models 
with OA and MHD EOS.
Almost all the models use the same nuclear reaction rates of Caughlan \&
Fowler. Model of CL, which uses OPAL opacity and tables of
MHD EOS, is close to our best models. There is a small difference in the
value of $\alpha$ (the difference in He mass fraction is due to different
age of the Sun). This difference stems possibly from their different heavy 
element abundance and low opacity table.
They use different initial mixture for heavy elements.
Model of BU includes the Coulomb interaction in EOS and the Los Alamos Opacity
(Cox et al. 1991).

The physical variables which determine the structure of the Sun is summarized 
in Table 2. These values (in $cgs$) are from the Model A which has the accuracy 
of $10^{-4}$ and is obtained by using MHD+GY and recent opacities. For 
accuracy of the model, number of shells is increased to $739$, 
during the evolution.  The energy that the Sun radiates is produced within the
inner $0.25~M_\odot$ core. Convective zone is between shells  256 and 727, where 
radiative temperature gradient exceeds adiabatic temperature gradient.
Its mass is only $2\%$ of the total mass of the Sun, and its distance 
to the center is $0.724 R_\odot$.
The total internal energy in outer shells is negative,
since binding energy of an atom having electronic configuration is considered negative.

Our conclusion is that solar models obtained by MHD EOS and OPAL
opacities are in closer agreement with the results
the helioseismology. The remaining small differences can be removed by 
taking into consideration the  diffusion process of He and heavy  elements.  
With MHD EOS and OPAL
opacity, and with diffusion process, Basu \& Thompson (BT) (1996) gives
the distance from bottom of the convective zone to the center as 
$0.714~R_{\odot}$. 
When one incorporates diffusion processes, 
Bahcall et al. (1995) emphasize that the decrease 
in mass fraction of He 
is about $10\%$
which gives better agreement with the observation. 
From Table 1, one sees that the main effect upon the initial He abundance
is provided by the adoption of the recent opacities which tend to
increase the abundance, while the MHD EOS has a smaller effect in the
opposite direction 

Structure of the Sun is not very sensitive to the methods of Henyey (Model
B) and Gabriel \& Y{\i}ld{\i}z (Model A) for the ionization of heavy
elements. But the method of Gabriel \& Y{\i}ld{\i}z is better than
Henyey method for rapid fitting processes, and, near the surface of the
Sun, it gives neutral heavy elements as expected.

\begin{acknowledgements}
We thank Dilhan Ezer-Eryurt and Maurice Gabriel for their valuable suggestions,
comments and discussion. 
\end{acknowledgements}

\def \apj#1#2{~ApJ~{#1}, #2}
\def \astroa#1#2{~astroa preprint}
\def \pr#1#2{~Phys.~Rev.~{#1}, #2}
\def \prt#1#2{~Phys.~Rep.~{#1}, #2}
\def \rmp#1#2{~Rev. Mod. Phys.~{#1}, #2}
\def \pt#1#2{~Phys.~Today.~{#1}, #2}
\def \pra#1#2{~Phys.~Rev.~{ A}~{#1}, #2}
\def \asap#1#2{~A\&A~{#1}, #2}
\def \asaps#1#2{~A\&AS~{#1}, #2}
\def \arasap#1#2{~Ann. Rev. Astron. Astrophys.~{#1}, #2}
\def \pf#1#2{~Phys.~~Fluids~{#1}, #2}
\def \apjs#1#2{~ApJS~{#1}, #2}
\def \pasj#1#2{~PASJ~{#1}, #2}
\def \mnras#1#2{~MNRAS~{#1}, #2}

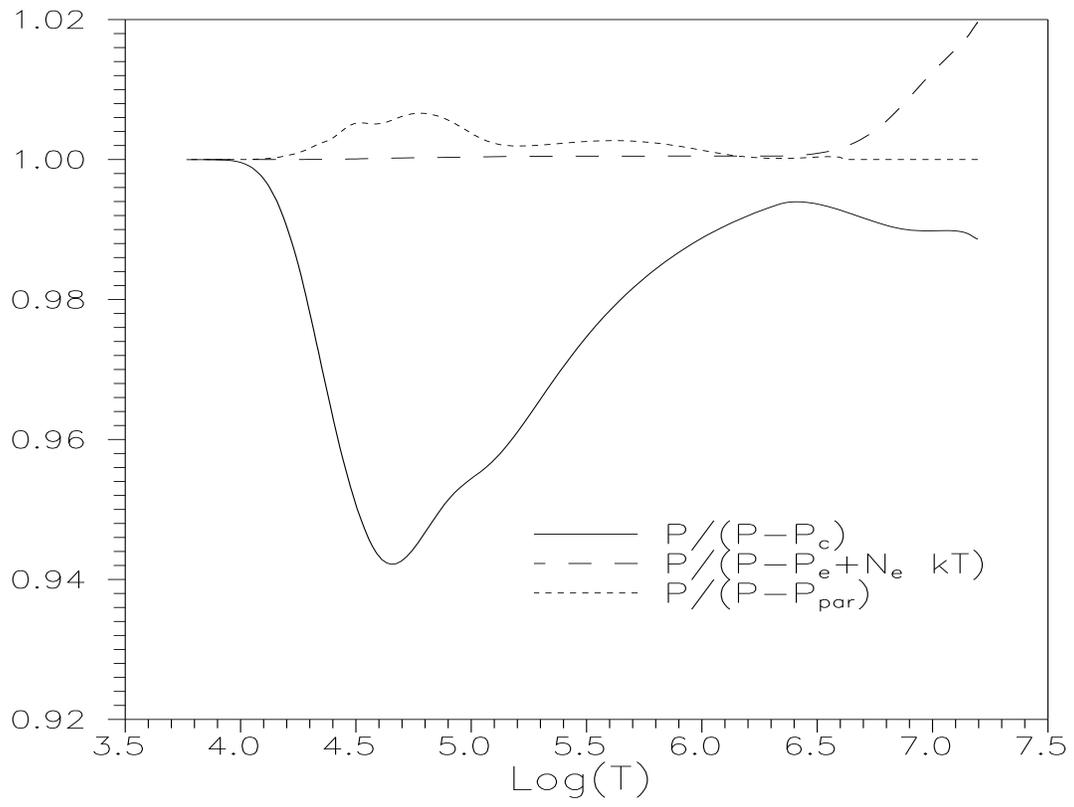
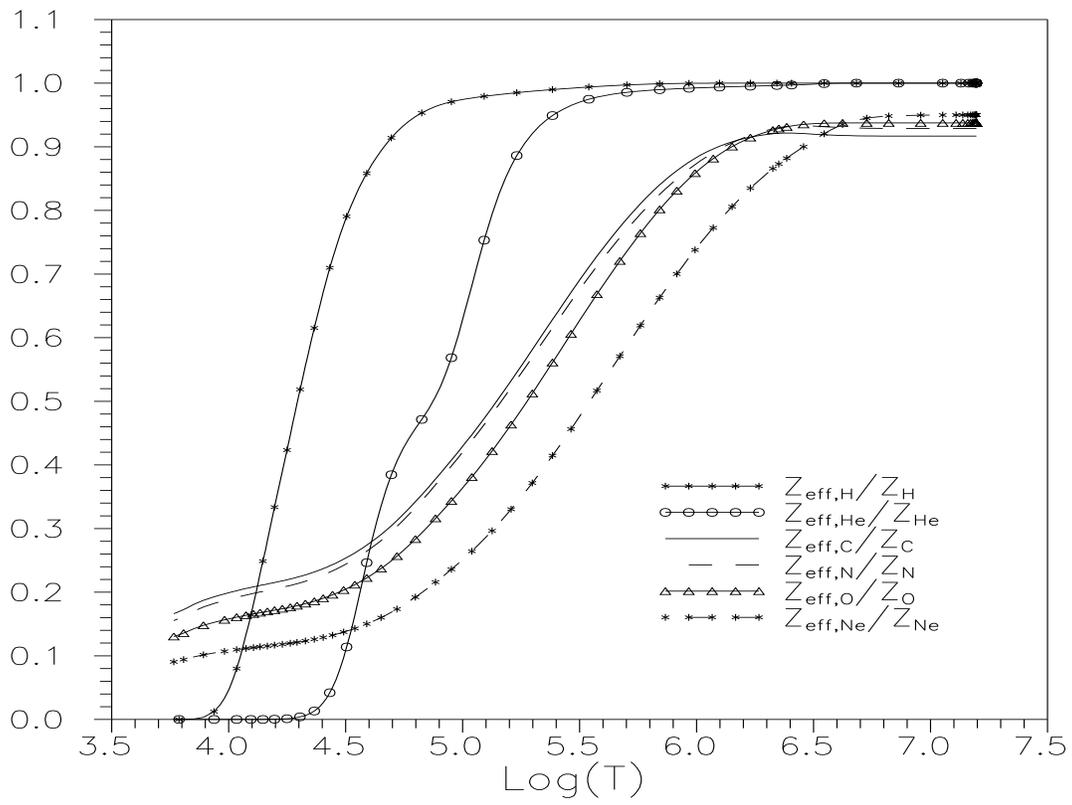
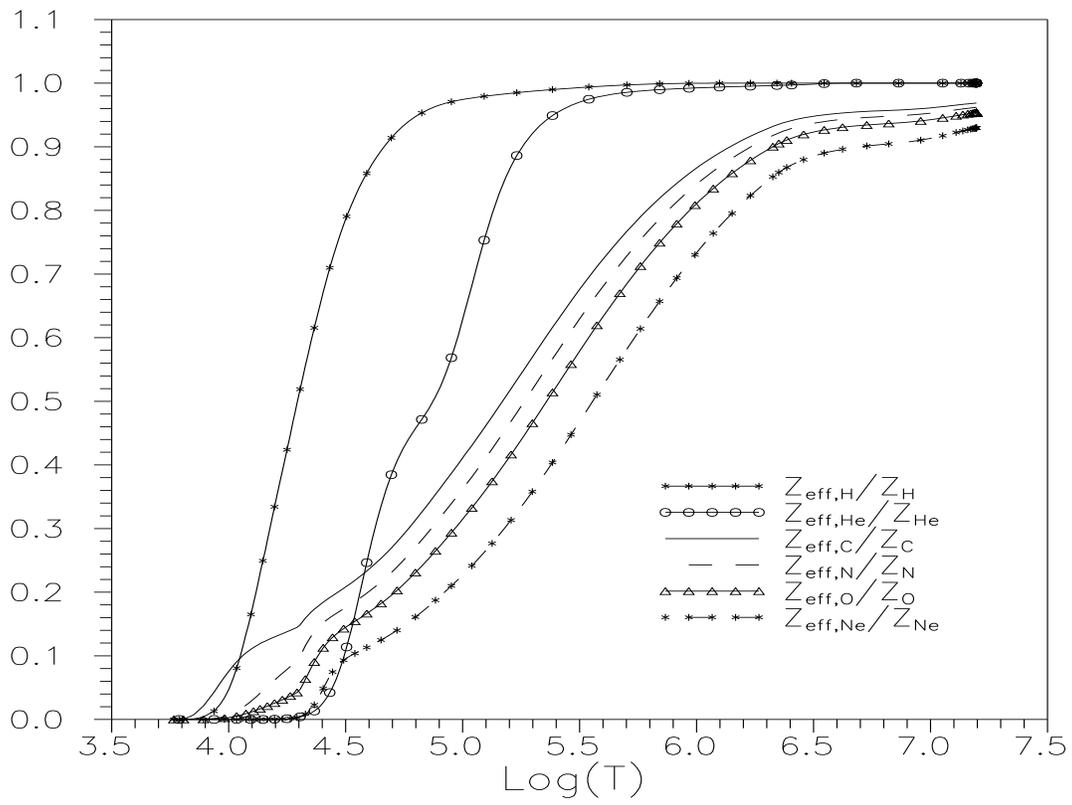
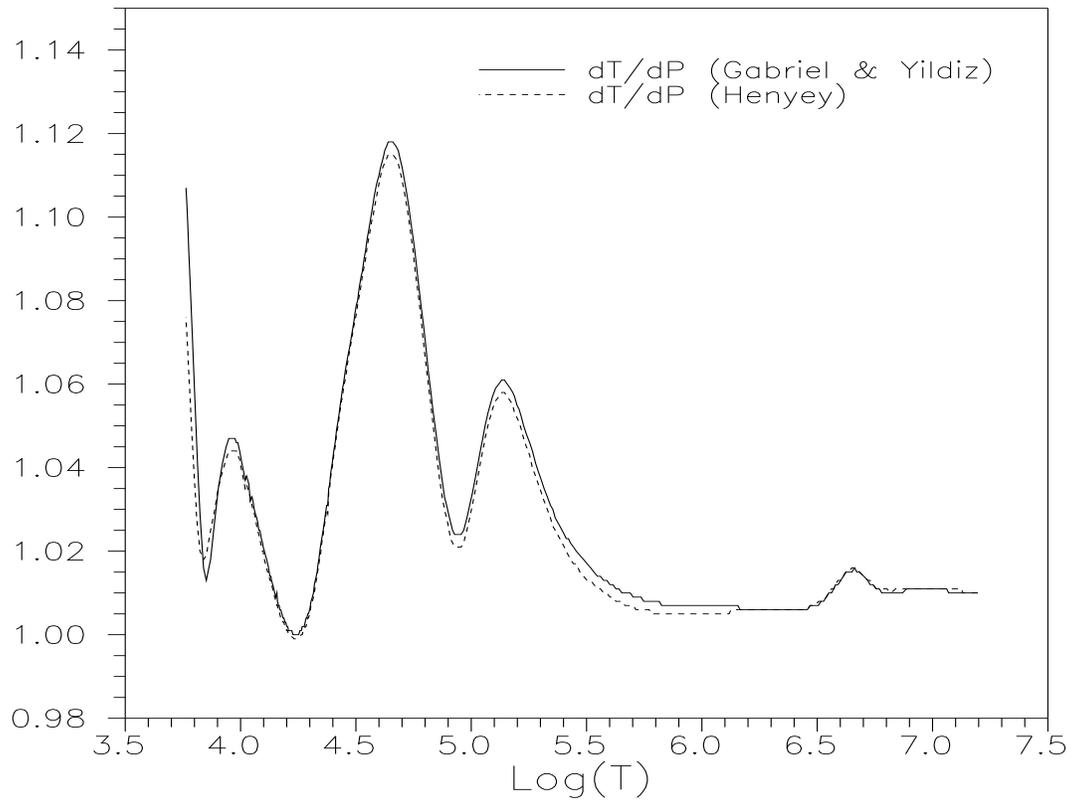
\end{document}